\documentclass[twocolumn,aps,prl,psfig,showpacs]{revtex4}
\usepackage{amsfonts}
\usepackage{amsmath}
\usepackage{graphicx}
\usepackage{bm}
\usepackage{amssymb}
\usepackage{times}
\usepackage{dcolumn}

\makeatletter

\begin{document}

\title{Half-Metallic Silicon Nanowires: Multiple Surface Dangling Bonds and
Nonmagnetic Doping}
\author{Zhuo Xu}
\author{Qing-Bo Yan}
\author{Qing-Rong Zheng}
\author{Gang Su}
\email[Corresponding author. ]{Email: gsu@gucas.ac.cn}
\affiliation{{College of Physical Sciences, Graduate University of Chinese Academy of
Sciences, P.O. Box 4588, Beijing 100049, China}}

\begin{abstract}
By means of first-principles density functional theory calculations,
we find that hydrogen-passivated ultrathin silicon nanowires (SiNWs)
along [100] direction with symmetrical multiple surface dangling
bonds (SDBs) and boron doping can have a half-metallic ground state
with 100\% spin polarization, where the half-metallicity is shown
quite robust against external electric fields. Under the
circumstances with various SDBs, the H-passivated SiNWs can also be
ferromagnetic or antiferromagnetic semiconductors. The present study
not only offers a possible route to engineer half-metallic SiNWs
without containing magnetic atoms but also sheds light on
manipulating spin-dependent properties of nanowires through surface
passivation.
\end{abstract}

\pacs{71.70.Ej, 73.21.Hb, 73.22.Dj}
\maketitle

It is widely believed that nanowires can be the most promising candidates
for the basic building blocks of future nanoelectronics, as they could be
essentially useful in engineering diverse nanodevices such as field-effect
transistors, logic gates, DNA sensors, etc. \cite%
{nanowire,device1,device2,device3} Thus, the nanowires with thin diameters
\cite{small1,small2} were actively studied both experimentally and
theoretically in the past years. Recently, particular attention was paid on
the ultrathin silicon nanowires (SiNWs) \cite%
{Chou,basic,doping,B+DB,001a,001b}, as they not only have unusual electronic
properties of scientific interest, but also could be potentially
incorporated into nowadays well-developed Si-based electronic technology.

The properties of ultrathin nanowires can be modified by adjusting diameters
and orientations, conducting the surface passivation, or introducing
dopants, etc. For SiNWs, among other things, it is known that (i) due to
surface reconstruction, SiNWs along [100] direction with diameters smaller
than 1.7 nm prefer a square cross section with sharp corners where the
electrons are quite localized \cite{001a,001b,shape,sharpcorner}; (ii) the
ultrathin SiNW that is saturated with H atoms on surface can show a
semiconducting behavior \cite{basic}, and if it is substitutionally doped
with nonmagnetic elements like B or P, the main bands of impurities
distribute at the top of valence bands \cite{B+DB}; (iii) H-passivated
ultrathin SiNWs along [100] direction interstitially doped by a certain
density of transition metals such as Co or Cr can have a half-metallic
ground state \cite{HM1}; (iv) when the H-passivated nanowires or surfaces
have an isolated or a single row of surface dangling bonds (SDBs), there is
a band contributed by the SDBs crossing the Fermi level (FL) \cite%
{B+DB,P+DB,BP+DB,SiGe+DB,C+DB}, and if the spin polarization is considered,
the band splits into two discrete bands, respectively, below and above the
FL \cite{C+DB}, leading to two band groups formed in the case of multiple
SDBs \cite{Si+DB}.

In the present Letter, by means of $ab$ $initio$ calculations we have found
that, for an ultrathin H-passivated SiNW along [100] direction with multiple
SDBs at symmetrical positions and substitutionally doping B atoms at the
center of cross section, a ferromagnetic (FM) configuration is favored as
the ground state, and such a SiNW is a half-metal with 100\% spin
polarization, namely, it is insulating for one spin direction while metallic
for the opposite spin direction at the FL. To the best of our knowledge,
this is the first time to report a novel type of half-metallic SiNWs with
multiple SDBs and nonmagnetic doping, in contrast to the previous study
where the half-metallic SiNWs were obtained by doping transition metals \cite%
{HM1}. This poses a possible route to make half-metallic SiNWs through
manipulating the surface passivation and introducing nonmagnetic dopants.

\begin{figure}[b]
\includegraphics[width=1.0\linewidth,clip]{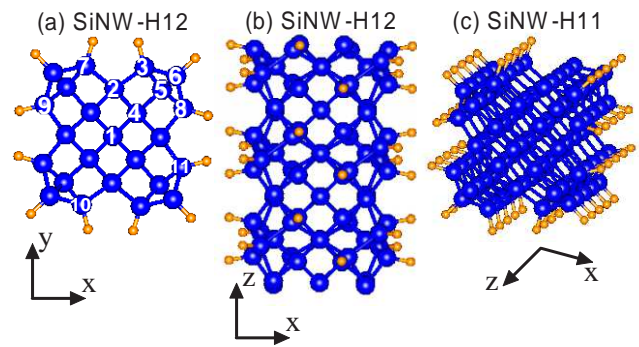}
\caption{(Color online) The cross section (a) and side view (b) of
the optimized structures of H-saturated SiNW along [100] direction.
The atomic positions marked by numbers in (a) are discussed in the
context. (c) The SiNW with one SDB per unit. Blue (dark gray) and
orange (light gray) balls represent Si and H atoms, respectively.}
\label{1}
\end{figure}

Our calculations were based on the density functional theory \cite{DFT} with
generalized gradient approximation expressed by PBE functional \cite{PBE},
employing norm-conserving pseudopotentials \cite{T-M} and linear
combinations of atomic orbitals. The SIESTA code \cite{SIESTA,ESPRESSO} with
double-zeta polarized basis sets was used to perform spin-polarized
calculations. The energy cutoff was 180 Ry, and the total energy convergence
criterion was 10$^{-5}$ eV. All structures were fully relaxed by the maximum
force tolerance of 0.015 eV/${\mathring{A}}$, and the lattice constant along
the wire axis was optimized for each structure. Among the SiNWs of 25 Si
atoms per unit [Fig. 1(a)] with different SDBs, the separation between two
adjacent wires was kept as 15.27 ($\pm $0.02) ${\mathring{A}}$. The
Brillouin zone along the axial direction was sampled with 15 k-points.

Figs. 1(a) and (b) depict the ultrathin SiNW along [100] direction
passivated with H atoms without SDBs (labeled by SiNW-H12), where there are
25 Si and 12 H atoms per unit cell. In SiNW-H12, the diameter is about 11 $%
\mathring{A}$ including H, and about 8 $\mathring{A}$ excluding H.
The positions in $y$ axis of the adjacent Si atoms at positions 3
and 4 [Fig. 1(a)] differ only by 0.009 $\mathring{A}$. Note that the
H saturated SiNW-H12 is semiconducting \cite{basic}. For the SiNW
with single SDB per unit [denoted by SiNW-H11, Fig. 1(c)], there is
one unpaired electron at the SDB in each unit, and the total
magnetic moment $\mu$ is 1.00 $\mu _{B}$ per unit cell ($\mu _{B}$
is Bohr magneton). In SiNW-H11, the Si atom with SDB [position 3 in
Fig. 1(a)] is stretched 0.047 $\mathring{A}$ outer than the Si at
position 4 that is still passivated with H, and the energy for
generating a single SDB per unit is 3.52 eV. We compared different
spin configurations by employing a supercell with two unit cells
(including two adjacent SDBs), and disclosed that the FM
configuration is favored as the ground state, with the energy of
11.3 meV per unit cell lower than the antiferromagnetic (AF)
configuration, and 155.6 meV per unit lower than the nonmagnetic
(NM) configuration. It turns out that the SiNW-H11 is a FM
semiconductor with an indirect gap of 0.59 eV.

\begin{figure}[tb]
\includegraphics[width=1.0\linewidth,clip]{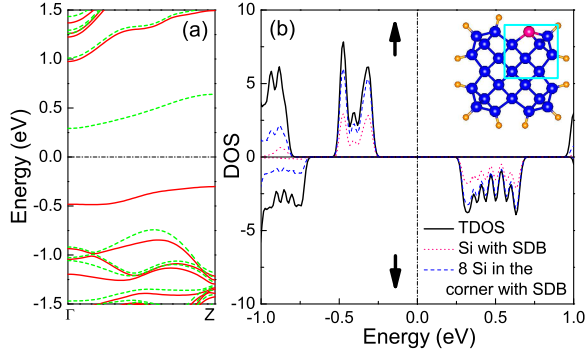}
\caption{(Color online) (a) The band structure and (b) the total and
projected density of states (DOS) of ferromagnetic SiNW-H11. The Si
with SDB is marked in pink (gray). The 8 Si atoms in the corner
including the SDB are enclosed in the inset of (b) and marked blue
(dark gray). The red (gray) solid and green (light gray) dashed
curves in band structures represent majority ($\uparrow$) and
minority ($\downarrow$) spins, respectively, throughout the context.
The Fermi energy is set to zero.} \label{2}
\end{figure}

The electronic structure and density of states (DOS) of FM SiNW-H11 are
presented in Fig. 2. It can be seen that there are two split bands just
above and below the FL [Fig. 2(a)]. The upper band is of minority spins ($%
\downarrow $), while the lower is of majority spins ($\uparrow $), which
correspond to the half-filled band crossing the FL if the spin polarization
is ignored, similar to the case at C(001) surface \cite{C+DB}. For the bands
away from the FL, there is merely a small displacement between the spin-up
and spin-down subbands, which is close to the SiNW-H12. The constituents of
the two split bands are clearly manifested in the projected DOS (PDOS): The
Si atom with a SDB contributes 44.9\% to the total DOS (TDOS), which
dominates overwhelmingly among all atoms, and the 8 Si atoms in the corner
including the SDB contribute 82.9\% to the TDOS, as shown in Fig. 2(b). This
indicates that the two split bands are formed primarily by the unpaired
electron of SDB and the electrons closely around the SDB.

\begin{figure}[tb]
\includegraphics[width=1.0\linewidth,clip]{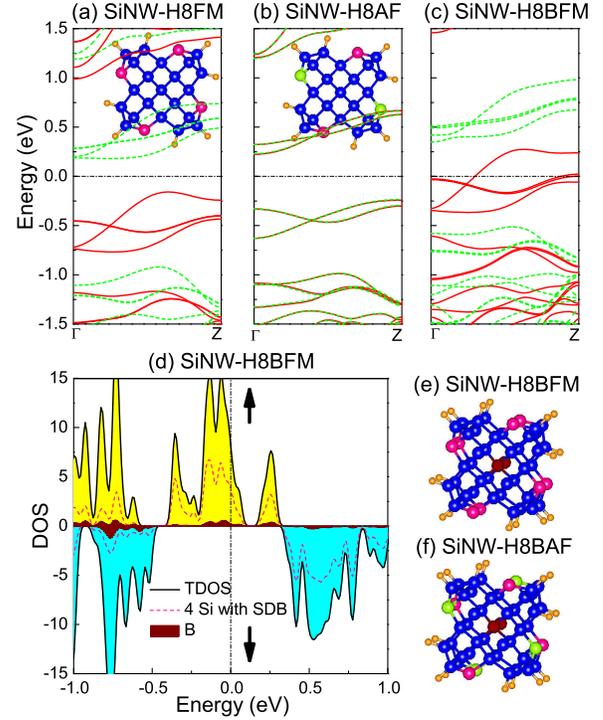}
\caption{(Color online) The band structures of (a) SiNW-H8FM, (b)
SiNW-H8AF, and (c) SiNW-H8BFM. (d) The DOS of SiNW-H8BFM. The
structure and spin configuration of SDBs for (e) SiNW-H8BFM and (f)
SiNW-H8BAF. The dark blue (dark gray), orange (light gray, small),
wine (solid) balls represent Si, H, B atoms, respectively; the pink
(gray) and green (light gray) atoms denote Si atoms with SDB for
majority and minority spins respectively.} \label{3}
\end{figure}

Now let us look at the SiNW with multiple SDBs. Without loss of
generality, the case of four SDBs per unit cell located
symmetrically on each edge of the cross section will be calculated,
as shown in Figs. 3(a) and (b), labeled by SiNW-H8. In comparison to
the SiNW-H11, the FM configuration of SiNW-H8 [SiNW-H8FM, Fig. 3(a)]
has more split bands, which are separated into two groups above and
below the FL, and each group includes four bands of the same spin
species (either spin up or down), where there is one band degenerate
with another in each group. The magnetic moment is 4.00 $\mu _{B}$
per unit cell. However, the SiNW-H8FM is merely a metastable state.
The ground state has an AF configuration [SiNW-H8AF, Fig. 3(b)],
where the SDBs on adjacent edges have opposite spin alignments, with
an energy lower than SiNW-H8FM by 26.9 meV per unit cell. The
average energy for generating each SDB in this case is 3.55 eV. The
spin-up and spin-down bands of the SiNW-H8AF coincide, and $\mu$ is
zero, as revealed in Fig. 3(b). Therefore, SiNW-H8 is an AF
semiconductor with an indirect gap of 0.47 eV. In the band structure
of SiNW-H8FM [Fig. 3(a)], the indirect gap between the bands of
majority and minority spins is 0.34 eV, which is so small that it
could be possible to generate a perfect half-metallic behavior in
the SiNW-H8 so long as the two conditions are satisfied: (i) The FL
is slightly shifted upward or downward to intersect the split bands
of the same spin species; (ii) The FM configuration of SDBs should
be the ground state. To meet with the conditions, it may be
convenient to conduct nonmetallic dopings (such as B or P).

By substitutionally doping one B atom at the center of the cross section
[position 1 in Fig. 1(a)] of the SiNW-H8FM (labeled by SiNW-H8BFM), we found
that the SiNW-H8BFM is the ground state, and the FL moves indeed downward
intersecting the majority spin bands, giving rise to a half-metallic
behavior, as shown in Fig. 3(c). This is also manifested in Fig. 3(d), where
the DOS of majority spin $G_{\uparrow}(E_{F})$ at the FL ($E_{F}$) is
8.28/eV, while the DOS of minority spin $G_{\downarrow }(E_{F})$ is zero,
leading to the spin polarization $P=[G_{\uparrow }(E_{F})-G_{\downarrow
}(E_{F})]/[G_{\uparrow }(E_{F})+G_{\downarrow }(E_{F})]=100\%$. The indirect
band gap of minority spin is 0.86 eV. The total magnetic moment of
SiNW-H8BFM is $3.00 $$\mu _{B}$ per unit cell, while the local moment of
each Si atom with SDB is 0.51 ($\pm $0.01) $\mu_{B}$, that is 0.11 $\mu _{B}$
lower than SiNW-H8AF. As there are three spin-up bands at different $k$
points in the Brillouin zone crossing the FL, it is hard to open a gap due
to Peierls instability \cite{001b}. Thus, the half-metallicity in the
SiNW-H8BFM is robust. We also examined possible magnetic configurations [FM,
AF and ferrimagnetic (FI)] in a supercell with either one or two unit cells,
respectively, together with NM (SiNW-H8BNM, spin-unpolarized) calculations,
and observed that apart from the FM ground state [SiNW-H8BFM, Fig. 3(e)],
the lowest metastable state is an AF metal [SiNW-H8BAF, Fig. 3(f)]. The
total energies of SiNW-H8BNM and SiNW-H8BAF are 309.2 meV and 21.5 meV per
unit cell, respectively, higher than that of SiNW-H8BFM. Based on the energy
difference between the SiNW-H8BFM and SiNW-H8BAF, we can estimate that at
the temperature lower than 250K, the half-metallic property in the SiNW-H8B
can be retained.

Doping P at the center in a FM configuration of SiNW-H8 (labeled by
SiNW-H8PFM) is similar to the SiNW-H8BFM except that the FL moves
upward crossing the spin-down bands, resulting in the
half-metallicity for minority spins. However, the ground state for
this case is a FI metal, where the total magnetic moment $\mu = 0.06
\mu_{B}$ per unit cell, and the total energy is slightly lower than
SiNW-H8PFM by 3.2 meV per unit cell. Compared with the AF ground
state of SiNW-H8, doping B at the center will decrease both the
charge of each SDB and the diameter of the SiNW, leading to a FM
ground state, while doping P brings about inverse changes, making a
FI ground state. Besides, it is unraveled that in spite of doping B,
P or not, if there are two SDBs in one corner [at positions 3 and 8
in Fig. 1(a)] or in one edge (at positions 3 and 7), the FM
configuration appears not to be the ground state.

\begin{figure}[tb]
\includegraphics[width=1.0\linewidth,clip]{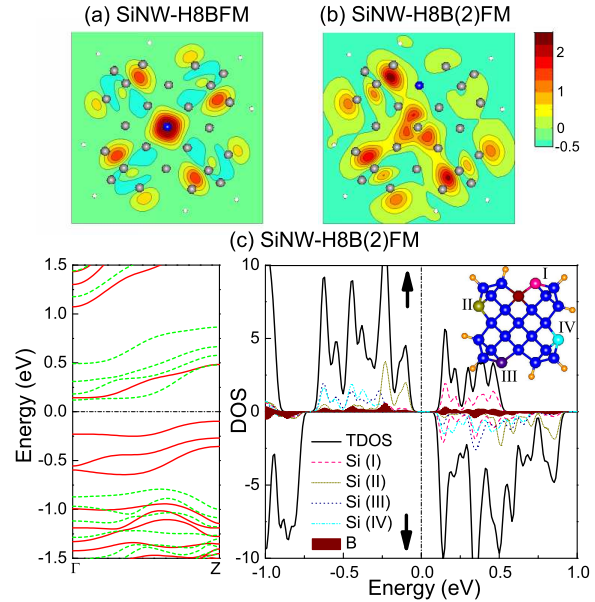}
\caption{(Color online) The contour plot of spin density
distribution (in unit of 10$^{-3}$ electrons/Bohr$^{3}$) of (a)
SiNW-H8BFM and (b) SiNW-H8B(2)FM at the cross section through the
central atom. The gray, white and dark blue (dark gray) balls stand
for the positions of Si, H and B atoms. (c) The band structure and
DOS of SiNW-H8B(2)FM. Four Si atoms with SDBs and one B atom are
marked pink (gray dash, I), dark yellow (gray short dot, II), purple
(dark gray dot, III), blue (light gray dash-dot, IV) and wine
(solid), respectively.} \label{4}
\end{figure}

For the SiNW-H8BFM, Fig. 4(a) indicates that the spin density ($\rho
_{\uparrow }-\rho _{\downarrow }$) distributes identically in each
of the four corners with SDB, where $\rho _{\uparrow }$ ($\rho
_{\downarrow }$) is the number density of electrons with spin up
(down). The dopant B at the center has a homogeneous effect on the
symmetrical SDBs, leading to the half-metallicity. If doping B takes
place at the edge [position 2 in Fig. 1(a)] instead of the center
with the FM configuration [SiNW-H8B(2)FM, Figs. 4(b) and (c)], there
are three spin-up bands just below the FL, and four spin-down bands
and one spin-up band just above the FL. The four Si atoms with SDBs
contribute to the PDOS distinctly, among which the one [red Si (I)
in Fig. 4(c)] adjacent to the B atom plays an overwhelming role. It
is also confirmed by the spin density distribution in Fig. 4(b),
where the spin density of the 1/4 cross section around Si (I) is
nearly zero, totally different from those in other three corners.
The local moment of Si (I), (II), (III) and (IV) is 0.09, 0.58, 0.63
and 0.61 $\mu _{B}$, respectively. In general, doping B at any
position other than the center will give a semiconductor. The total
energy per unit cell of doping B atoms at position 4, 2, 3, 5, 6 and
7 [Fig. 1(a)] is 0.36, 0.52, 0.70, 0.46, 0.58 and 0.73 eV,
respectively, higher than doping at position 1 (SiNW-H8BFM),
suggesting the center is the most preferred doping position for B.

For the FM configuration, one B dopant per unit will reduce one
electron below the FL, which, depending on the relative positions
among dopants and SDBs, could change the band structure in two ways.
One is that the dopant affects mainly one position of SDBs, moving
one spin-up band to above the FL, as SiNW-H8B(2)FM. The other is
that the dopant affects equally the multiple positions of SDBs,
pushing more spin-up bands crossing the FL, as the half-metallic
SiNW-H8BFM. Consequently, in order to produce the half-metallicity,
the SDBs should be arranged at symmetrical positions and, the
dopants must be located symmetrically to the SDBs. In addition,
several cases with larger radius of 61 Si atoms and four symmetrical
SDBs per unit cell are also checked. The results show that
substitutionally doping either one B atom at the center or five
around the center per unit cell can still generate the FM ground
state and lead to the half-metallicity with 100\% spin polarization.

\begin{figure}[tb]
\includegraphics[width=1.0\linewidth,clip]{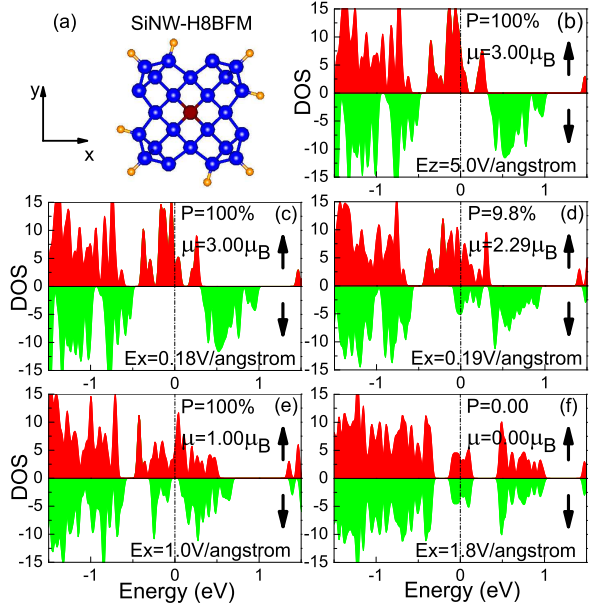}
\caption{(Color online) (a) The SiNW-H8BFM under an external electric field $%
\vec{E}$. The DOS under (b) $E_{z}$ = 5.0 V/$\mathring{A}$ along $z$
direction; and (c) $E_{x}$ = 0.18, (d) 0.19, (e) 1.0 and (f) 1.8 V/$%
\mathring{A}$ along $x$ direction.} \label{5}
\end{figure}

To examine whether the half-metallic property of the SiNW-H8BFM [Fig. 5(a)]
is stable under external electric fields, we made calculations in presence
of longitudinal and transverse electric fields, respectively. By applying an
electric field along the nanowire direction (along $z$ axis), we found that
the half-metallic behavior of the SiNW-H8BFM still retains under a field as
large as 5.0 $V/\mathring{A}$ [Fig. 5(b)]. When an electric field is applied
perpendicular to the nanowire axis (along $x$ direction), we observed that
the half-metallic state is kept when the field is less than 0.18 $V/%
\mathring{A}$ [Fig. 5(c)]; when the transverse field reaches 0.19 $V/%
\mathring{A}$ [Fig. 5(d)] or larger, the spin polarization decreases
dramatically, and the half-metallic property is destroyed, which becomes a
FM metal. When the electric field increases further, say about 1.0 $V/%
\mathring{A}$, the half-metallic property comes back again [Fig. 5(e)]. When
the field exceeds 1.8 $V/\mathring{A}$ [Fig. 5(f)], it becomes a NM metal.

In summary, a novel type of half-metallic SiNWs along [100] direction by
introducing multiple SDBs and boron doping is predicted by means of the
spin-dependent \textit{ab initio} calculations. The obtained results show
that the half-metallicity in such SiNWs is quite robust against external
electric fields. Under the circumstances with different SDBs, the
H-passivated SiNWs can also be FM or AF semiconductors. The present findings
might be applicable in nanospintronics and nanomagnetism.

The authors are grateful to X. Chen, S. S. Gong, X. L. Sheng, Z. C.
Wang, and L. Z. Zhang for helpful discussions. The calculations are
performed on the supercomputer NOVASCALE 6800 in Supercomputing
Center of CAS. This work is supported in part by the NSFC (Grant No.
10625419), the MOST of China (Grant No. 2006CB601102), and CAS.


\begin{thebibliography}{99}
\bibitem{nanowire} J. Appenzeller, J. Knoch, M. T. Bj\"{o}rk, H. Riel, H.
Schmid, and W. Riess, IEEE Trans. Electron Devices \textbf{55}, 2827 (2008).

\bibitem{device1} D. Wang, B. A. Sheriff, and J. R. Heath, Nano Lett.
\textbf{6}, 1096 (2006).

\bibitem{device2} C. M. Lieber \textit{et al.}, Science \textbf{294}, 1313
(2001).

\bibitem{device3} Guo-Jun Zhang \textit{et al.}, Nano Lett. \textbf{8}, 1066
(2008).

\bibitem{small1} A. M. Morales and C. M. Lieber, Science \textbf{279}, 208
(1998).

\bibitem{small2} D. D. D. Ma \textit{et al.}, Science \textbf{299}, 1874 (2003).

\bibitem{Chou} J. A. Yan, L. Yang, and M. Y. Chou, Phys. Rev. B \textbf{76},
115319 (2007).

\bibitem{basic} E. Durgun, N. Akman, C. Ataca, and S. Ciraci, Phys. Rev. B
\textbf{76}, 245323 (2007).

\bibitem{doping} A. K. Singh, V. Kumar, R. Note, and Y. Kawazoe, Nano Lett.
\textbf{6}, 920 (2006).

\bibitem{B+DB} C. R. Leao, A. Fazzio, and A. J. R. da Silva, Nano Lett.
\textbf{8}, 1866 (2008).

\bibitem{001a} R. Rurali and N. Lorente, Phys. Rev. Lett. \textbf{94},
026805 (2005).

\bibitem{001b} R. Rurali, Phys. Rev. B \textbf{71}, 205405 (2005).

\bibitem{shape} S. Ismail-Beigi and T. Arias, Phys. Rev. B \textbf{57},
11923 (1998).

\bibitem{sharpcorner} J. X. Cao, X. G. Gong, J. X. Zhong, and R. Q. Wu,
Phys. Rev. Lett. \textbf{97}, 136105 (2006).

\bibitem{HM1} E. Durgun, D. Cakir, N. Akman, and S. Ciraci, Phys. Rev. Lett.
\textbf{99}, 256806 (2007); E. Durgun, N. Akman, and S. Ciraci,
Phys. Rev. B \textbf{78}, 195116 (2008).

\bibitem{P+DB} M.-V. Fern\'{a}ndez-Serra, Ch. Adessi, and X. Blase, Nano
Lett. \textbf{6}, 2674 (2006).

\bibitem{BP+DB} H. Peelaers, B. Partoens, and F. M. Peeters, Nano Lett.
\textbf{6}, 2781 (2006).

\bibitem{SiGe+DB} R. Kagimura, R. W. Nunes, and H. Chacham, Phys. Rev. Lett.
\textbf{98}, 026801 (2007).

\bibitem{C+DB} J. Cho and J. Choi, Phys. Rev. B \textbf{77}, 075404 (2008).

\bibitem{Si+DB} S. Okada, K. Shiraishi, and A. Oshiyama, Phys. Rev. Lett.
\textbf{90}, 026803 (2003).

\bibitem{DFT} W. Kohn and L. J. Sham, Phys. Rev. \textbf{140}, A1133 (1965).

\bibitem{PBE} J. P. Perdew, K. Burke, and M. Ernzerhof, Phys. Rev. Lett.
\textbf{77}, 3865 (1996).

\bibitem{T-M} N. Troullier and J. L. Martins, Phys. Rev. B \textbf{43}, 1993
(1991).

\bibitem{SIESTA} P. Ordej\'{o}n, E. Artacho, J. M. Soler, Phys. Rev. B
\textbf{53}, 10441 (1996).

\bibitem{ESPRESSO} Main properties have been confirmed in supercells of two
and four units along z axis by SIESTA. We also used Quantum ESPESSO
(P. Giannozzi \emph{et al.}, www.quantum-espresso.org) with plane
wave basis sets and ultrasoft pseudopotentials to verify the
relative energies of different spin configurations of SiNW-H8B in a
unit cell, which gives similar results with SIESTA.

\end{thebibliography}
\end{document}